\documentclass[11.5 pt, A4size, journal]{IEEEtran}

\usepackage{amsmath,amsfonts}
\usepackage{xcolor}
\usepackage{algorithmic}
\usepackage{array}
\usepackage[caption=false,font=normalsize,labelfont=sf,textfont=sf]{subfig}
\usepackage{textcomp}
\usepackage{stfloats}
\usepackage{url}
\usepackage{verbatim}
\usepackage{graphicx}
\usepackage{multirow} 
\def\BibTeX{{\rm B\kern-.05em{\sc i\kern-.025em b}\kern-.08em
    T\kern-.1667em\lower.7ex\hbox{E}\kern-.125emX}}
\usepackage{balance}
\usepackage{setspace}
\begin{document}
\title{CNN-Based Automated Parameter Extraction Framework for Modeling Memristive Devices}

\author{Akif Hamid and Orchi Hassan*\\
\textit{Department of Electrical and Electronic Engineering, \\ Bangladesh University of Engineering and Technology, Dhaka -1205, Bangladesh.}\\
*email: orchi@eee.buet.ac.bd
\vspace {-0.5 cm}}

\maketitle

\begin{abstract}
Resistive random access memory (RRAM) is a promising candidate for next-generation nonvolatile memory (NVM) and in-memory computing applications. Compact models are essential for analyzing the circuit and system-level performance of experimental RRAM devices. However, most existing RRAM compact models rely on multiple fitting parameters to reproduce the device I–V characteristics, and in most cases, as the parameters are not directly related to measurable quantities, their extraction requires extensive manual tuning, making the process time-consuming and limiting adaptability across different devices. This work presents an automated framework for extracting the fitting parameters of the widely used Stanford RRAM model directly from the device I-V characteristics. The framework employs a convolutional neural network (CNN) trained on a synthetic dataset to generate initial parameter estimates, which are then refined through three heuristic optimization blocks that minimize errors via adaptive binary search in the parameter space. We evaluated the framework using four key NVM metrics: set voltage, reset voltage, hysteresis loop area, and low resistance state (LRS) slope. Benchmarking against RRAM device characteristics derived from previously reported Stanford model fits, other analytical models, and experimental data shows that the framework achieves low error across diverse device characteristics, offering a fast, reliable, and robust solution for RRAM modeling.
\end{abstract}


\section{Introduction}
Memristors are emerging electronic devices whose internal states depend on the history of the applied current or voltage\cite{mem_1,mem_2}. They can encode information through distinct resistance states and, when integrated into crossbar arrays, have great potential for applications in both data storage and artificial intelligence \cite{IMC,IMC_2,IMC_3,AI_Cross_1,AI_Cross_2,AI_Cross_3,AI_Cross_4,AI_Cross_5}. The structural simplicity, compact-size, multibit storage, and intrinsic compatibility with conventional CMOS fabrication processes make them a strong candidate for next generation non volatile memory (NVM) \cite{memory_2,MRS_1,MRS_2,MRS_3} applications.

Memristive devices featuring a metal-insulator-metal (MIM) configuration, where an oxide layer is positioned between the top and bottom electrodes, are commonly designated resistive random access memory (RRAM) in the literature \cite{RRAM,Stanford,VerilogCompact,Stanford_Spice}, a convention that will be maintained throughout this work. Their resistive switching behavior arises from the formation and dissolution of conductive filaments within the oxide layer, typically represented by an internal state variable $w$, and driven by ionic and vacancy migration \cite{ionic,ionic_2,ionic_3}. Accurate representation of these switching dynamics is essential. Circuit designers rely on SPICE or Verilog-A models \cite{spice_1,spice_2,spice_3,verilog_2,VerilogCompact} to design and evaluate system architectures, while device researchers primarily report experimental I–V characteristics. Compact models translate experimental findings into circuit simulations. The resistive switching phenomenon has been captured in numerous analytical compact models \cite{modelling,modelling_2,modelling_3}, which are generally built on two fundamental equations:
\begin{align}
\frac{dw}{dt} &= f_1(w,V) \\
I &= f_2(w,V)
\end{align}
where $w$ represents the state variable, V is applied voltage, and I is device current. The coupling between current and state-variable equations induces path-dependent behavior, making conventional regression-based techniques for parameter extraction ineffective for RRAM. Moreover, most compact models rely on multiple fitting parameters that lack direct correlation with standard RRAM metrics, making it difficult to reproduce experimental data accurately. Consequently, manual parameter tuning remains the most widely adopted approach for parameter extraction in RRAM compact models \cite{Yakopcic},\cite{Manual},\cite{Stanford}. The procedure heavily relies on domain expertise and systematic parameter adjustment, as the interdependent parameters require multiple iterations and careful selection of initial values by the practitioner. Particle Swarm Optimization (PSO) technique offers an algorithmic alternative \cite{PSO} but suffers from initialization sensitivity, local minima convergence, and poor scalability in high-dimensional parameter spaces. Physics-informed machine learning approaches eliminate need for manual parameter extraction, but they often suffer from poor generalization due to overfitting and incur high computational overhead that degrades large-scale simulation performance \cite{ML1},\cite{ML2}. 

\begin{figure}[!t]
    \centering
    \includegraphics[width=0.48\textwidth]{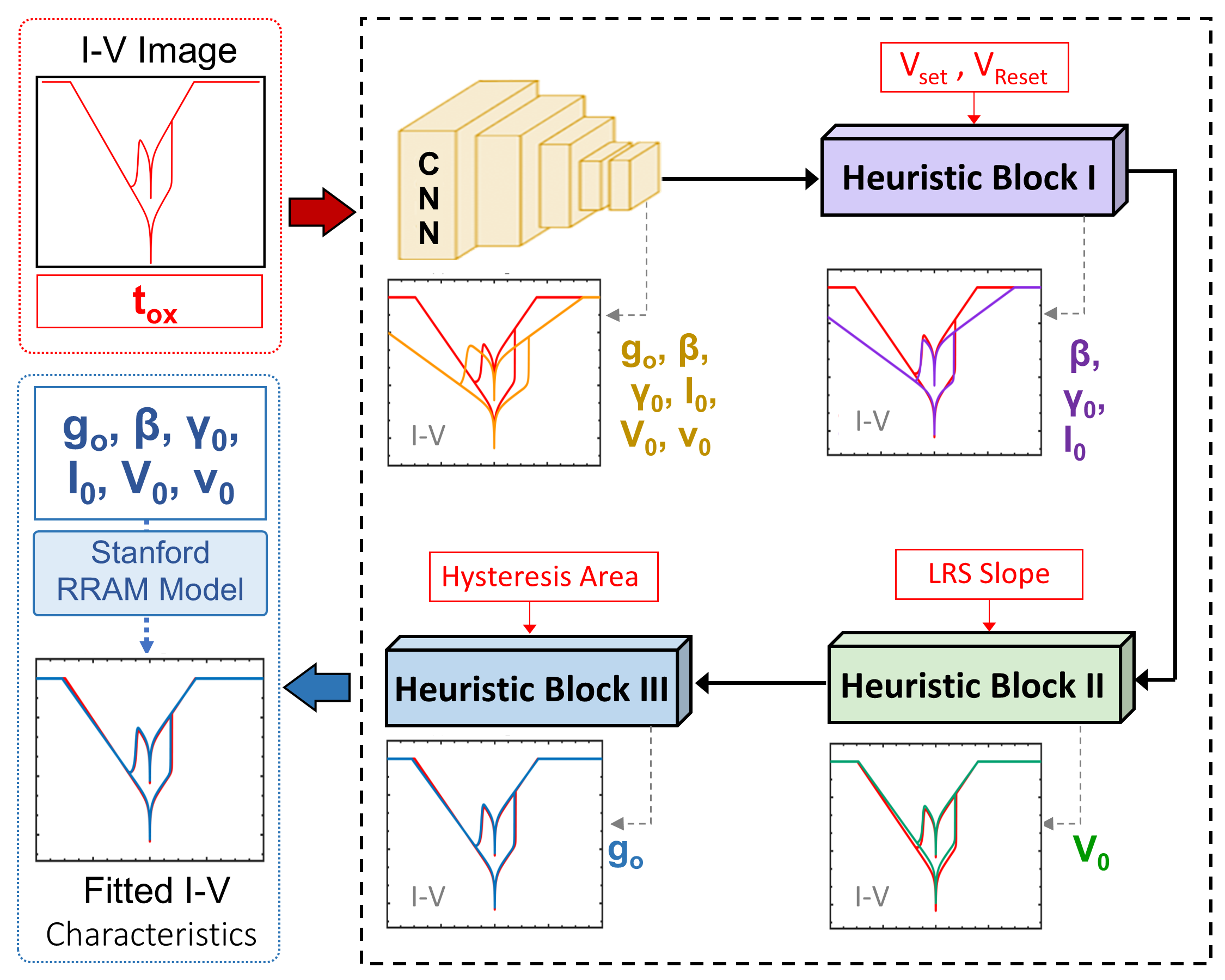}
    \caption{Proposed automated parameter extraction framework for Stanford RRAM model consists of a CNN block followed by three sequential heuristic blocks. The CNN block processes an RGB image of RRAM I-V characteristics along with oxide thickness to generate an initial parameter fit. The first heuristic block optimizes parameters $\beta$ and $\gamma_{0}$ to match set/reset reference voltages and scales the I-V curve. The second heuristic block adjusts parameter $V_0$ to match the reference LRS slope. The third heuristic block tunes parameter $g_0$ to match the reference hysteresis area. The framework outputs optimized Stanford RRAM model fitting parameters after the complete processing sequence.}
    \label{fig:framework}
\end{figure}

This work presents an automated fitting parameter extraction framework for the Stanford RRAM model, which takes in device I-V characteristics in image format, the oxide thickness, $t_{ox}$ and four directly extractable key NVM reference metrics: set/reset voltage, hysteresis area, and LRS slope, to predict the required fitting parameters as illustrated in Fig.~\ref{fig:framework}. The framework employs a convolutional neural network (CNN) block to make the initial parameter estimates, and subsequent heuristics blocks fine-tune the initial parameters to match the discussed NVM reference metrics. Our framework is designed to extract the parameters of the Stanford RRAM model which is described in Section II. The detailed description of the proposed automated framework and its performance evaluation are presented in Section III and IV, respectively.


\section{Stanford RRAM Model}
The Stanford RRAM Model simplifies ion and vacancy migration to a single dominant filament while maintaining essential switching physics. This model centers on the tunneling gap (g) between the filament tip and electrode, with switching dynamics described by equations (1-3). 

\begin{align}
I &= I_0 \times \exp\left(-\frac{g}{g_0}\right) \times \sinh\left(\frac{V}{V_0}\right) \tag{1} \\
\frac{dg}{dt} &= -v_0 \times \exp\left(-\frac{E_a}{kT}\right) \times \sinh\left(\frac{\gamma \times a_0}{t_{ox}} \times \frac{qV}{kT}\right) \tag{2} \\
\gamma &= \gamma_0 - \beta \cdot g^3 \tag{3} 
\end{align}

Current conduction depends exponentially on the tunneling gap distance (1), which is calculated by considering electric field effects, temperature-enhanced oxygen ion migration (2-3). Key parameters include applied voltage (V), current (I), activation energy ($E_a$), atomic spacing ($a_o$), oxide thickness ($t_{ox}$), and various fitting parameters ($I_0, g_0, V_0, v_0, \beta, \gamma_0$) that enable device-specific calibration. It should be noted that the correlation between these parameters and NVM metrics is not straightforward; even when some parameters show correlation, as illustrated in Fig.~\ref{fig:Variations}, altering other interdependent parameters can produce a completely different fit. For the purposes of this analysis, the local temperature variations arising from Joule heating effects and stochastic effects present in the model have been excluded.

\section{Proposed Automated Framework for Parameter Tuning}
The proposed framework utilizes the I–V image of an RRAM device, $t_{ox}$, and four NVM metrics derived from device I-V characteristics to generate the fitting parameters. It comprises of a convolutional neural network (CNN) block that generates initial parameter estimates, effectively replicating the domain expertise typically required for initial parameter estimation. This approach is well-founded, as the CNN is trained on an extensive dataset comprising thousands of RRAM I-V characteristics images, enabling it to learn the underlying patterns and relationships inherent in RRAM behavior.

However, relying solely on CNN-based parameter estimation presents significant limitations, particularly the tendency toward overfitting. Such an approach would demonstrate poor generalization performance when applied to I-V characteristics sourced from external literature, thereby restricting the framework's applicability. To mitigate this limitation and enhance the framework's generalization capabilities, we introduce three additional heuristic blocks within our framework. These blocks employ adaptive binary search algorithms to iteratively refine selected parameters, ensuring convergence toward optimal values based on predefined NVM evaluation metrics (detailed in subsequent sections). This hybrid approach combines the pattern recognition capabilities of deep learning with the systematic optimization strengths of heuristic blocks, resulting in a generalizable parameter fitting framework.

\subsection{Dataset Preparation}
\begin{figure}[!htb]
    \centering
    \includegraphics[width=0.48\textwidth]{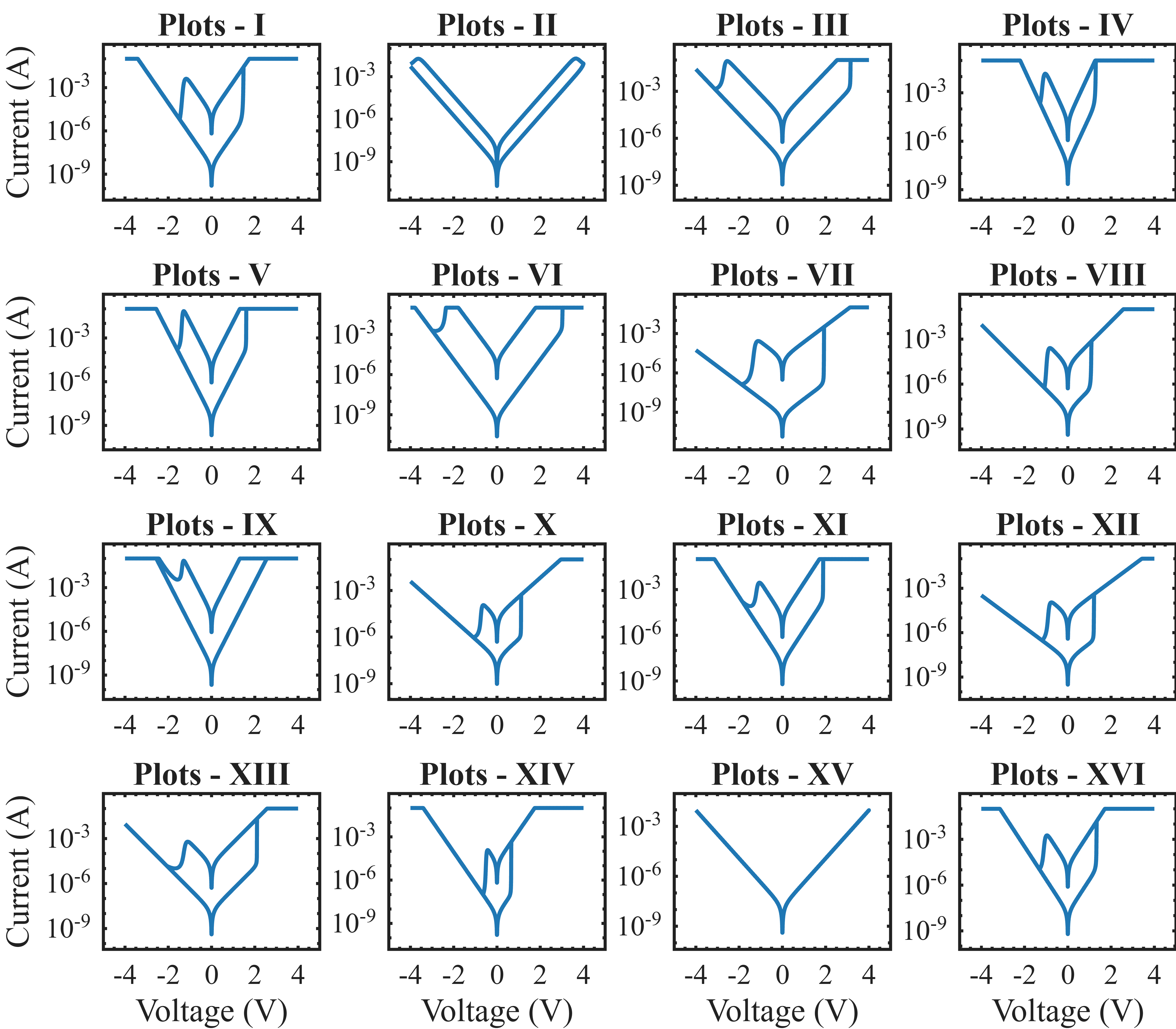}
    \caption{Representative synthetic I-V curves for RRAM devices under different parameter configurations. The dataset was generated through systematic perturbation of fitting parameters.}
     \label{fig:Samples}
\end{figure}

To develop the automated framework for the Stanford RRAM model, a synthetic dataset comprising approximately 16,000 images was generated in Python. These images were produced by varying five fitting parameters $g_0$, $V_0$, $\nu_{\text{0}}$, $\beta$, and $\gamma_0$ along with $t_{\text{ox}}$, ensuring a broad representation of possible I-V characteristics. $I_0$, was held constant at $100\,\mu A$ since it is a scaling factor and does not directly influence the shape of the I-V characteristics. Material parameters such as $a_{o}$ and $E_{a}$ for the Stanford RRAM model were adopted from \cite{VerilogCompact} and the system was solved using Euler method as cited in the work. Each image was of dimensions $224\times224\times3$ and stored in RGB format, enabling the use of transfer learning in later stages of the model development process. Voltage range between -4V and 4V was selected, ensuring that most I-V curves exhibited hysteresis behavior. A compliance current of $100~ \rm mA$  was imposed to prevent the current from exceeding realistic physical limits. Representative examples from the synthetic dataset are shown in Fig.~\ref{fig:Samples}. I-V curves which did not exhibit hysteresis or lacked clearly defined set and reset voltages were pruned from the dataset. 

\subsection{Convolutional Neural Network Block}
To emulate experience in the fitting parameter extraction process a CNN block is incorporated. For the CNN block, a ResNetV2 (50) architecture \cite{Resnet} was employed. Transfer learning was leveraged by  freezing first 20 layers of the ResNetV2 to reduce the training time and utilize high frequency feature extraction capabilities from the earlier layers.

\begin{figure}[!b]
    \centering
    \includegraphics[width=0.47\textwidth]{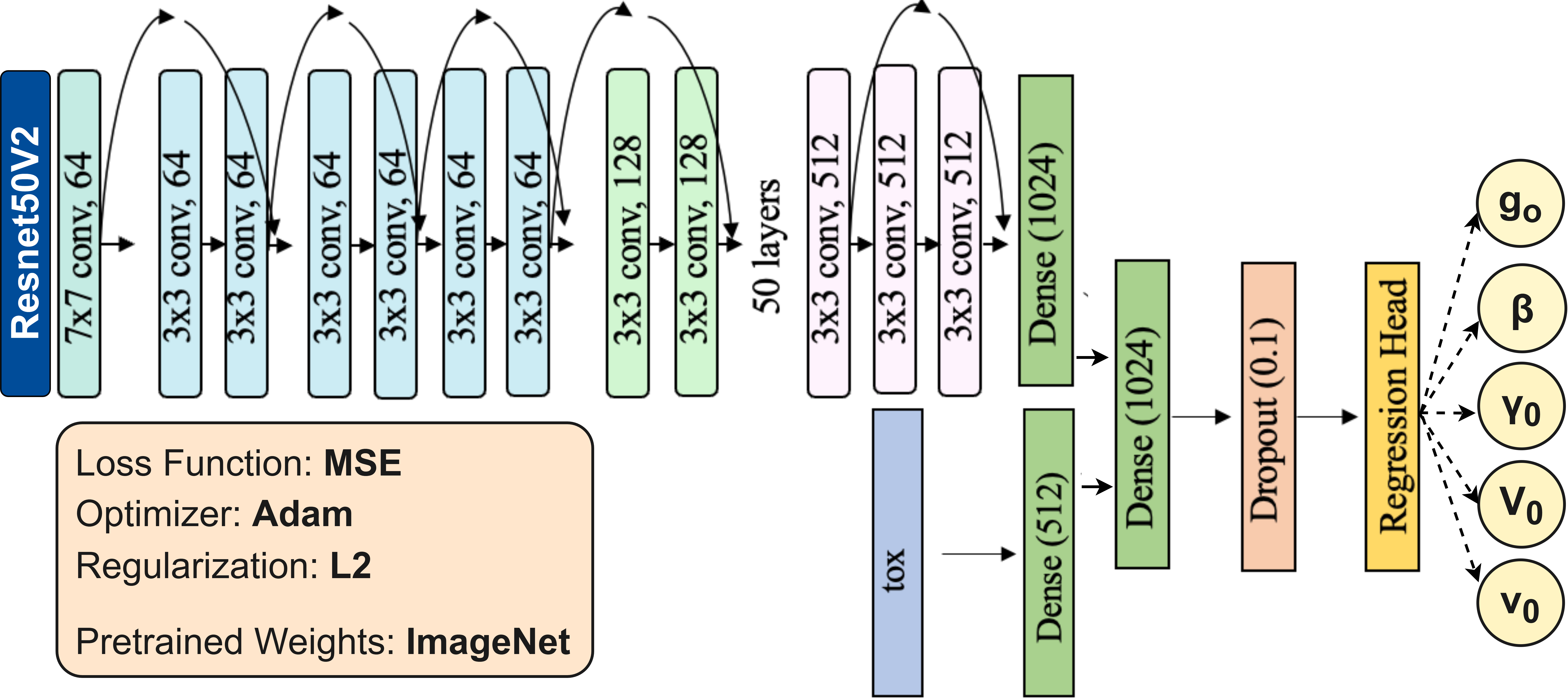}
    \caption{Convolutional neural network architecture for initial parameter estimation. The model employs a pretrained ResNet-50v2 backbone, with modifications to the CNN block for RRAM parameter fitting.}
     \label{fig:Model}
\end{figure}

The classification head of the original ResNetV2 model was replaced with a custom regression head designed to output five parameters rather than six. Notably, the parameter $I_0$ was excluded from being predicted by the convolutional neural network (CNN) due to the lack of inherent scale in the input images. Instead, $I_0$ was determined post hoc using heuristic blocks. Along with the images of I-V characteristics, the oxide thickness ($t_{\text{ox}}$) was provided as an auxiliary input. This value passed through a separate dense layer and was concatenated with the final output of the ResNetV2 model before being fed into the regression head as shown in Fig.~\ref{fig:Model}. This structure allowed the model to account for variations in oxide thickness, which significantly impact the I-V characteristics of RRAM devices. 

An Adam optimizer was used to train the model, with an initial learning rate set to $10^{-3}$. A learning rate scheduler was employed, decaying the learning rate exponentially by a factor of $e^{-0.1x}$ every epoch after first 10 epochs. Mean squared error (MSE) was chosen as the loss function for this regression task. The model was trained for a total of 54 epochs before early stopping was triggered based on validation loss improvement. The final root mean squared error (RMSE) after training was $0.1094$ on validation set. At inference, the I-V characteristics is preprocessed into the correct image format ($224\times224\times3$) using Matplotlibs default plotting scheme before passing into the CNN block.

\subsection{Evaluation Metrics}
\begin{figure}[!t]
    \centering
    \includegraphics[width=0.48\textwidth]{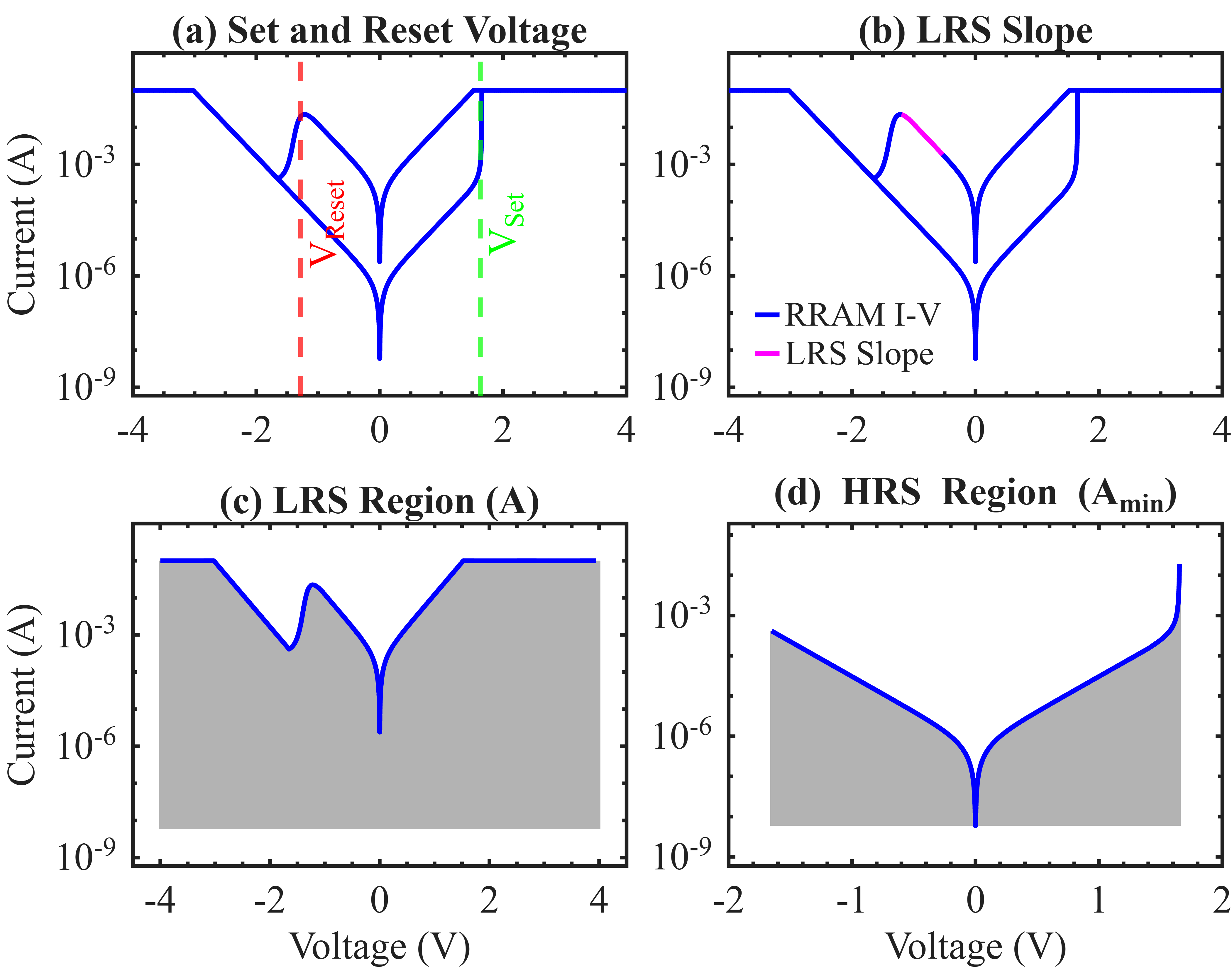}
    \caption{Key metrics extraction from device I-V curve: (a) Highlights $V_{\text{set}}$ and $V_{\text{reset}}$ (b) LRS slope in the reset region, (c) Area enclosed by the LRS region, and (d) Area enclosed by the HRS region.}
    \label{fig:metric_extraction}
\end{figure}

To assess the performance of the automated framework in accurately reproducing target I-V characteristics for non-volatile memory applications, custom evaluation metrics are proposed as illustrated in Fig.~\ref{fig:metric_extraction}. Rather than employing conventional mean squared error (MSE), which computes point-to-point differences across the entire curve, these application-specific metrics were designed to evaluate the framework's performance within the context of non-volatile memory requirements. Conventional error metrics such as MSE prioritize certain device artifacts that are significant for neuromorphic computing or in-memory computing applications, which the underlying Stanford RRAM model itself cannot accurately represent in some cases. To avoid such limitations and adequately capture the performance aspects most relevant NVM application such custom metrics are proposed.

\subsection{Heuristic Blocks}
To refine the initial fitting parameters generated by the CNN block and achieve better alignment with target evaluation metrics, three sequential heuristic blocks are employed. Each block is designed to calibrate specific fitting parameters through iterative adaptive binary search algorithm, thereby minimizing discrepancies between desired and fitted evaluation metrics.

\begin{figure}[!t]
    \centering
    \includegraphics[width=0.48\textwidth]{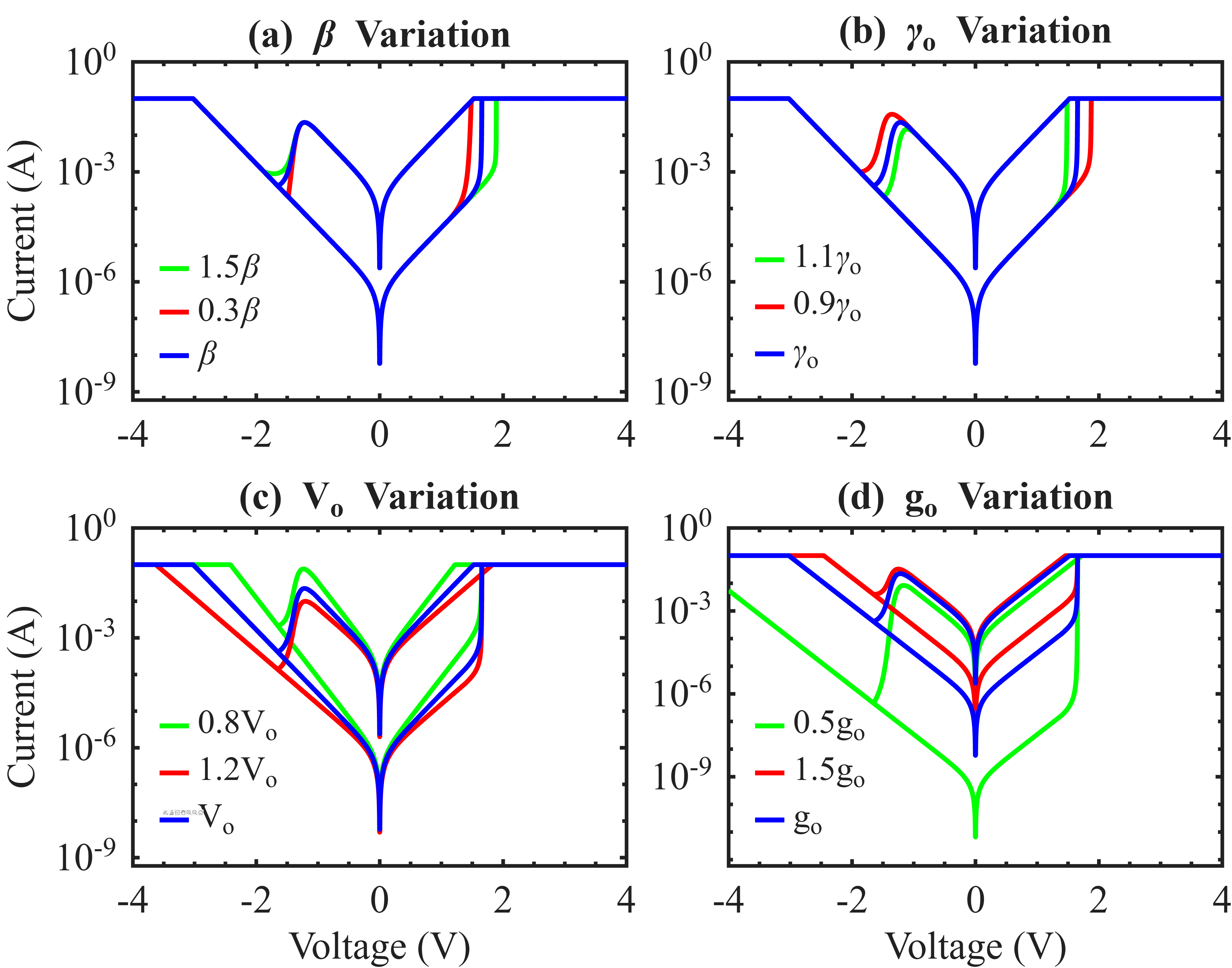}
    \caption{Effect of fitting parameter variation on RRAM I-V Characteristics: (a) Increasing $\gamma_0$ reduces the reset voltage($V_{\text{reset}}$), (b) increasing $\beta$ elevates the set voltage ($V_{\text{set}}$), (c) $V_0$ modulates the switching dynamics as reflected in the LRS slope, and (d) $g_0$ controls the hysteresis loop area ($A$ and $A_{\text{min}}$).}
    \label{fig:Variations}
\end{figure}

The parameter fitting strategy is informed by established relationships between model parameters and device characteristics, as documented in previous studies \cite{Manual} and illustrated in Fig.~\ref{fig:Variations}. Specifically, increasing $\gamma_0$ primarily reduces the reset voltage ($V_{\text{reset}}$), while increasing $\beta$ elevates the set voltage ($V_{\text{set}}$). The parameter $V_0$ modulates switching dynamics and influences the low-resistance state (LRS) slope, whereas $g_0$ affects the hysteresis area characteristics ($A$ and $A_{\text{min}}$). Each heuristic block requires user-specified reference values for the evaluation metric being optimized as indicated in Fig.~\ref{fig:framework}. Corresponding algorithmic implementations have been developed to systematically extract these evaluation metrics from simulated I-V characteristics.

\begin{figure}[!t]
    \centering
    \includegraphics[width=0.48\textwidth]{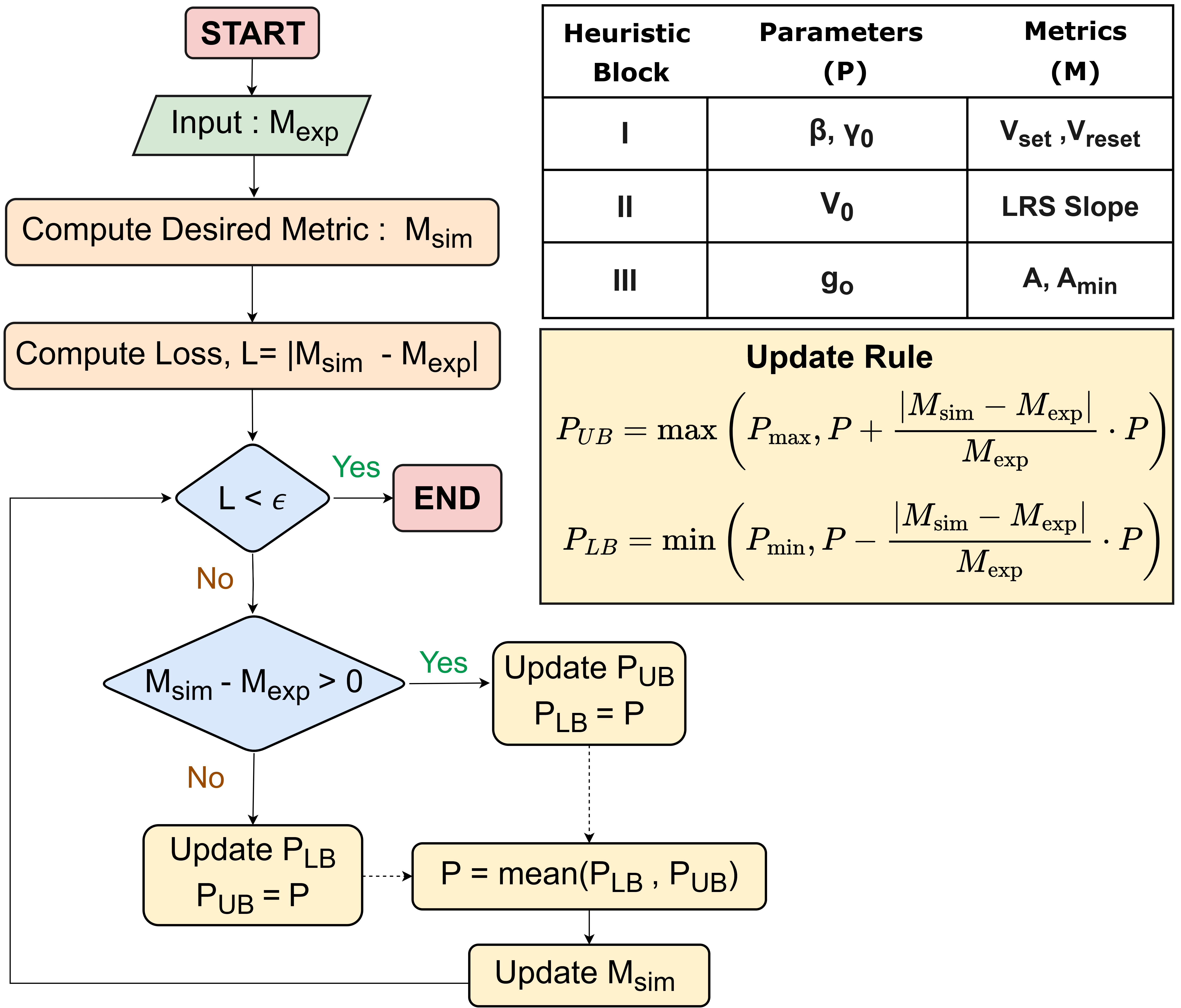}
    \caption{Adaptive binary search algorithm employed in each heuristic block to effectively match evaluation metrics. Each heuristic block employs this algorithm to match target evaluation metrics by optimizing $\gamma_0$ for $V_{\text{reset}}$, $\beta$ for $V_{\text{set}}$, $V_{\text{o}}$ for LRS slope, and $g_{\text{o}}$ for hysteresis area.}
    \label{fig:Heuristics}
\end{figure}

The initial heuristic block focuses on calibrating $\gamma_0$ and $\beta$ to achieve accurate set and reset voltage matching. The parameter $\gamma_0$ is adjusted to align the modeled reset voltage with the desired $V_{\text{reset}}$, while $\beta$ is tuned to ensure correspondence between modeled and desired $V_{\text{set}}$ values. The loss function for this optimization block is formulated as the sum of absolute differences between modeled and desired values of $V_{\text{set}}$ and $V_{\text{reset}}$, accounting for the coupled effects of $\gamma_0$ and $\beta$ on both voltage parameters. 

The second heuristic block involves tuning the parameter $V_0$, which governs the slopes of both high-resistance state (HRS) and LRS regions within the I-V characteristic. The heuristic block initially attempts to extract the LRS slope near the set region ($V_{set}$); however, if this region is flattened due to compliance current, it adaptively prioritizes slope estimation in the reset region ($V_{reset}$) to ensure accurate slope reproduction. 

The final heuristic block adjusts the parameter $g_0$ to optimize the hysteresis area, ensuring alignment with desired I-V characteristics. However, LRS-enclosed area typically exceeds the HRS-enclosed area by several orders of magnitude, a fundamental characteristic of RRAM devices. To prevent the LRS area from overpowering the HRS area, the heuristic block prioritizes matching the areas of both the LRS and HRS regions ($A$ and $A_{min}$), thereby maintaining reasonable correspondence with device I–V characteristics. For numerical stability, parameter ranges were constrained in the framework as $\gamma_0 \in [0, 24],\; \beta \in [0, 2.1],\; V_0 \in [0.15, 0.4],\; g_0 \in [1.5 \times 10^{-10},2.5 \times 10^{-10}]$.

\section{Results of fitting RRAM IV Characteristics Using Proposed Framework}
In this section, we evaluate the performance of the proposed framework by fitting the I-V characteristics of various RRAM devices with the Stanford RRAM model, where the required fitting parameters are automatically extracted using our framework. The framework was evaluated on both seen and unseen test cases across three categories of I–V characteristics:

\begin{enumerate}
    \item \textbf{Stanford Model Benchmarking:} Reproducing I-V characteristics that were previously fitted using the Stanford RRAM model, serving as a benchmark for comparison.
    \item \textbf{Fit with other Analytical Models:} Modeling I-V characteristics derived from other analytical models, such as VTEAM and Yakopcic with the Stanford RRAM model.
    \item \textbf{Fit with Experimental Data:} Experimentally reported I-V characteristics obtained directly from the literature which were reproduced using the Stanford RRAM model.
\end{enumerate}
Benchmarking across the three categories was conducted to assess the framework’s robustness to device characteristics beyond its training scope, thereby validating its adaptability. To analyze the quality of the produced fit using extracted parameters, we evaluated relative error percentage in the discussed NVM metrics.

\subsection{Stanford Model Benchmarking:}  
\begin{figure}[!t]
    \centering
   \includegraphics[width=0.48\textwidth]{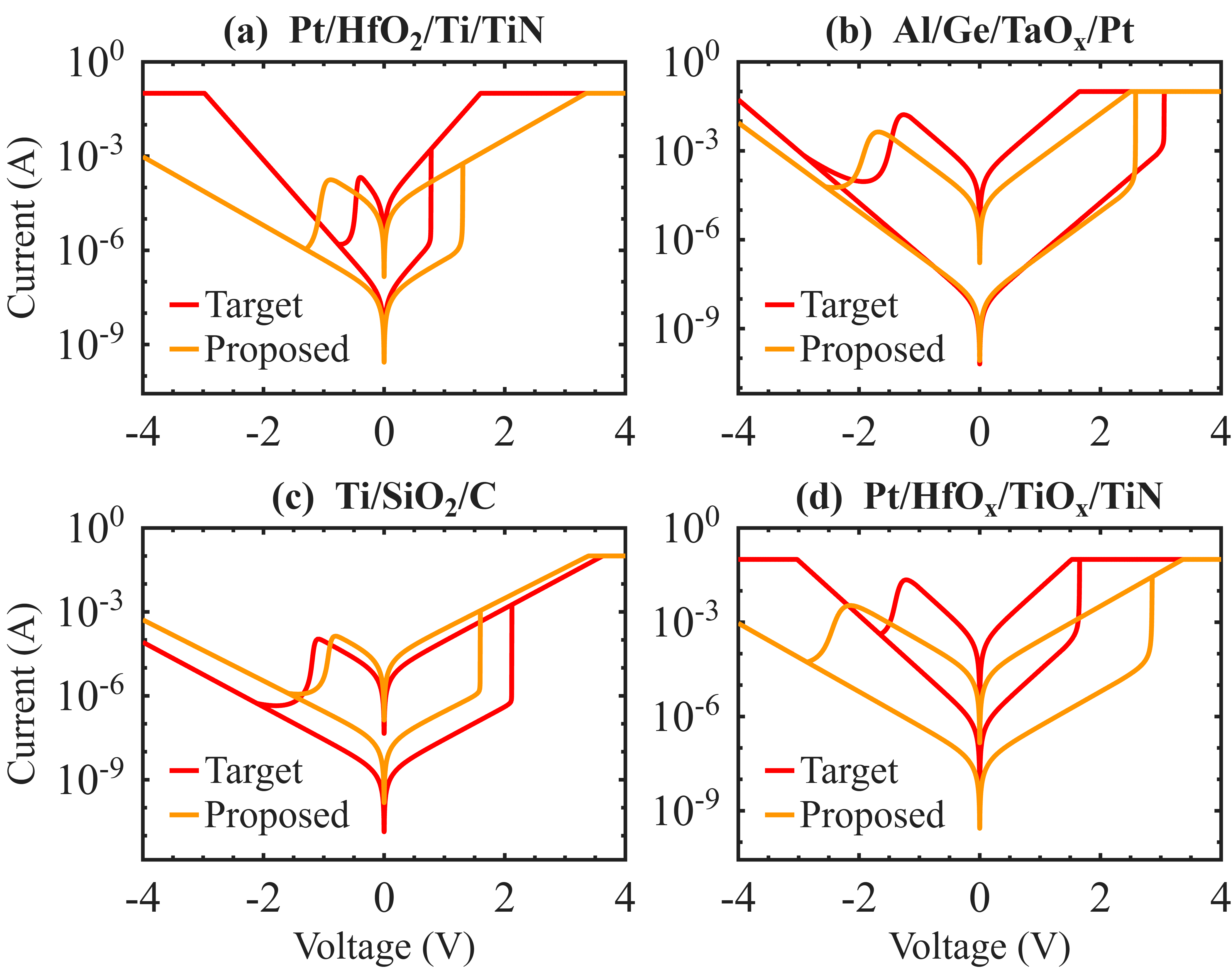}
    \caption{Fitting results from the CNN block initial estimates (yellow) with previously fitted data (red) for four different device characteristics from \cite{Manual} are shown. The extracted parameters for each device are (a) $g_0 = 2.39 \times 10^{-10}$, $V_0 = 0.397$, $\nu_{\text{0}} = 10.20$, $I_0 = 1\times 10^{-4}$, $\gamma_0 = 8.87$, $\beta = 0.504$; 
(b) $g_0 = 1.97 \times 10^{-10}$, $V_0 = 0.290$, $\nu_{\text{0}} = 9.98$, $I_0 = 1 \times 10^{-4}$, $\gamma_0 = 9.44$, $\beta = 0.686$; 
(c) $g_0 = 2.21 \times 10^{-10}$, $V_0 = 0.398$, $\nu_{\text{0}} = 0.0127$, $I_0 = 1 \times 10^{-4}$, $\gamma_0 = 14.0$, $\beta = 1.43$; and 
(d) $g_0 = 2.39 \times 10^{-10}$, $V_0 = 0.399$, $\nu_{\text{0}} = 2.27$, $I_0 = 1 \times 10^{-4}$, $\gamma_0 = 9.67$, $\beta = 0.463$..}
     \label{fig:comparison}

\end{figure}

\begin{figure}[!t]
    \centering
    
      \includegraphics[width=0.48\textwidth]{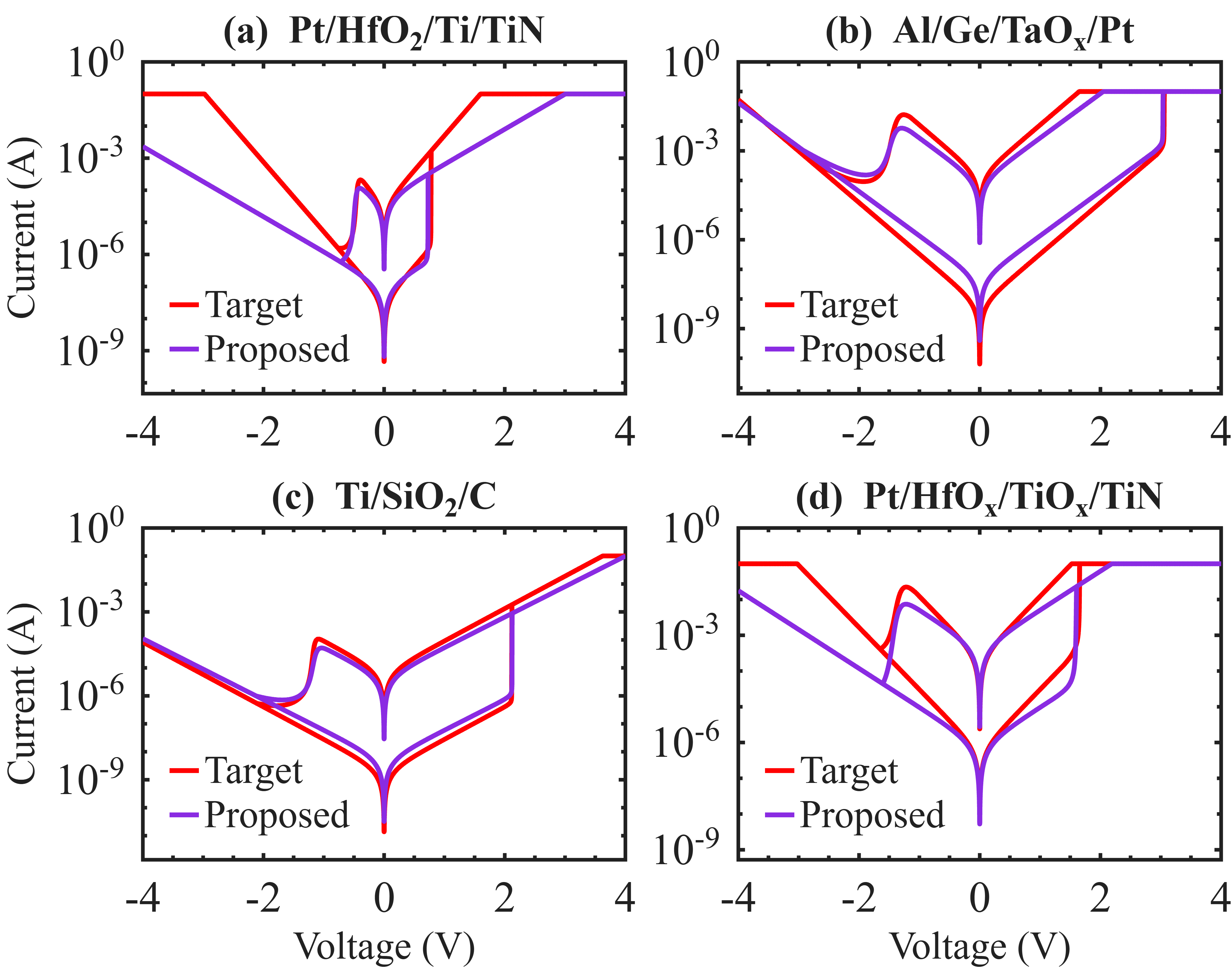}
  
    \caption{Fitting results after heuristic Block-I (purple) shows improved matching of set and reset voltages. Optimized parameters from Block-I are 
    (a) $I_0 = 2.4 \times 10^{-4}$, $\gamma_0 = 1.41$, $\beta = 1.69$; 
    (b) $I_0 = 4.7 \times 10^{-4}$, $\gamma_0 = 1.34$, $\beta = 1.51$; 
    (c) $I_0 = 2.12 \times 10^{-5}$, $\gamma_0 = 10.85$, $\beta = 1.16$; and 
    (d) $I_0 = 1.9 \times 10^{-3}$, $\gamma_0 = 0.551$, $\beta = 0.603$.}
    \label{fig:heuristic1}
\end{figure}

\begin{figure}[!t]
    \centering
    \includegraphics[width=0.48\textwidth]{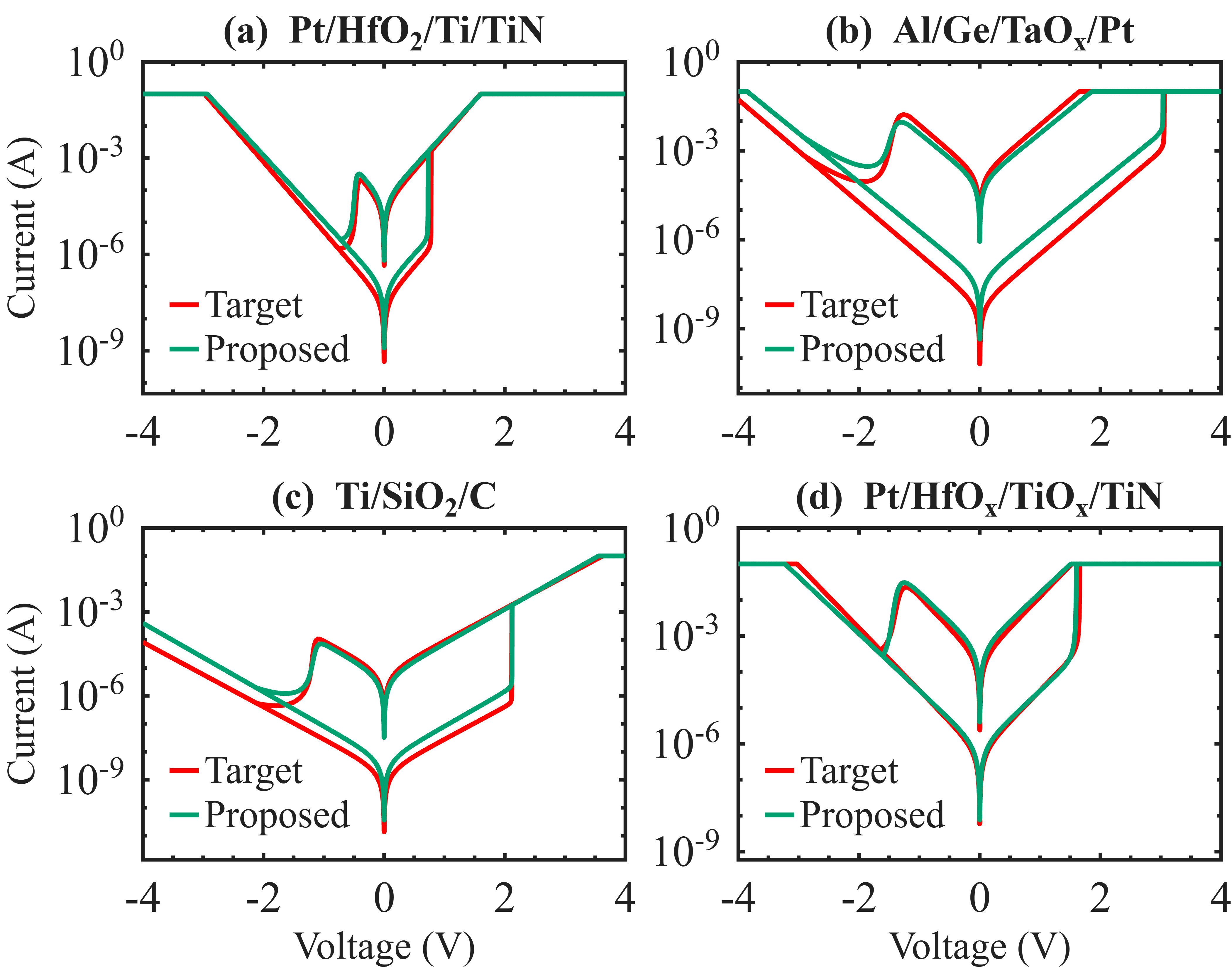}
    \caption{Fitting results after heuristic Block-II (green) showing further improved matching of HRS and LRS slope. Optimized $V_0$ values are (a) $0.208$, (b) $0.261$, (c) $0.355$, and (d) $0.275$.}
    \label{fig:heuristic2}
\end{figure}

\begin{figure}[!t]
    \centering
  \includegraphics[width=0.48\textwidth]{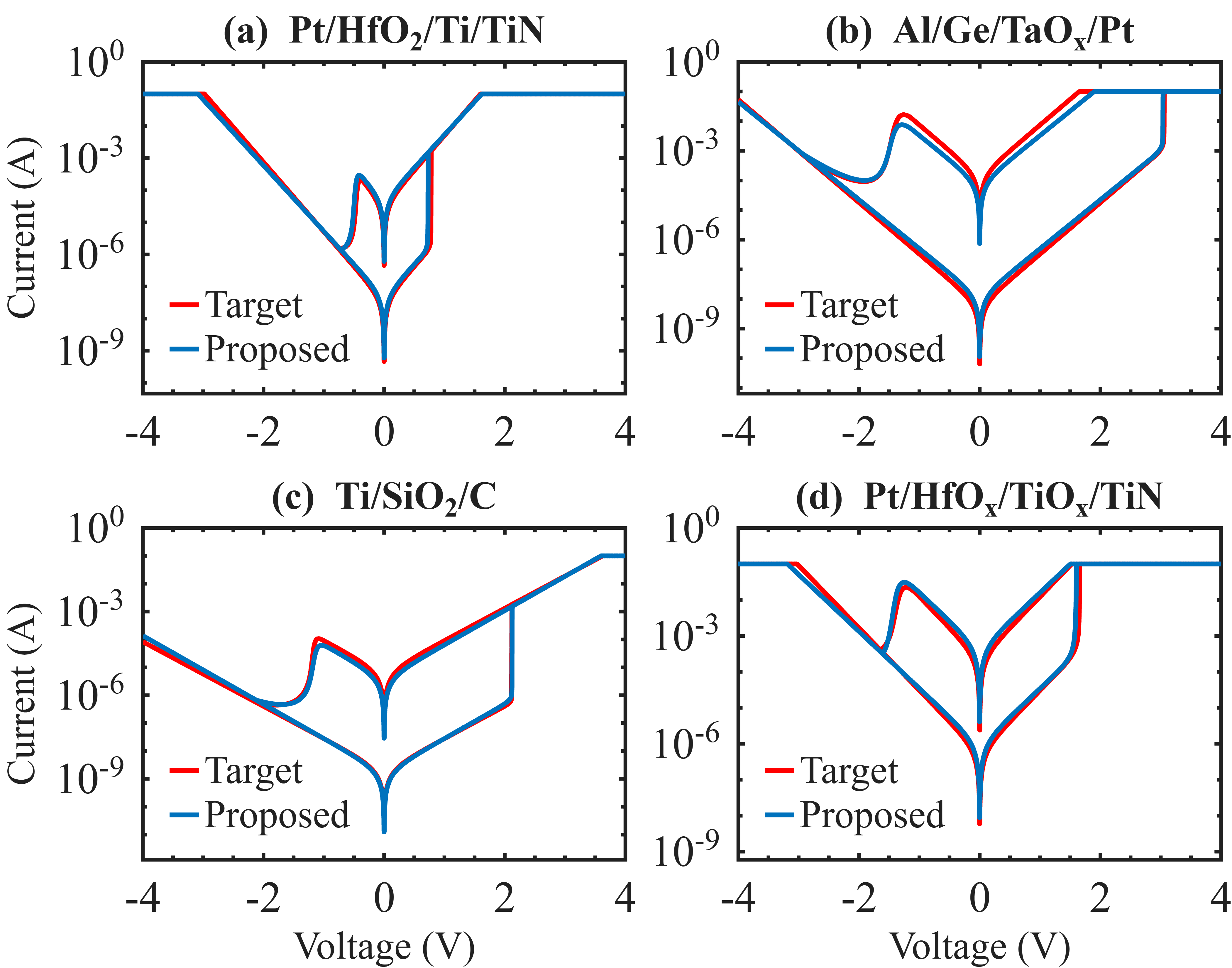}
    \caption{Fitting results after final heuristic Block-III, showing improved matching of the area enclosed by the hysteresis loop. Optimized $g_0$ values are (a) $2.17 \times 10^{-10}$, (b) $1.65 \times 10^{-10}$, (c) $1.94 \times 10^{-10}$, and (d) $2.44 \times 10^{-10}$.
}
    \label{fig:heuristic3}
\end{figure}

Firstly, we evaluate the results of our fitting parameter extraction process against four device I–V characteristics previously fitted with the Stanford RRAM model~\cite{Manual}, as shown in Fig.~\ref{fig:comparison}, \ref{fig:heuristic1}, \ref{fig:heuristic2}, and \ref{fig:heuristic3}. Image of I-V characteristics and $t_{ox}$, as specified in \cite{Manual} were passed into the CNN block. Key parameters such as $V_\text{set}$, $V_\text{reset}$, hysteresis area, and LRS slope were extracted manually as per Fig.~\ref{fig:metric_extraction} and passed into the heuristic blocks.

\begin{table}[!t]
    \centering
    \scriptsize
    \setlength{\tabcolsep}{1.5pt} 
    \caption{Reference NVM metrics for RRAM devices previously fitted using Stanford RRAM model.}
    \label{tab:rram_nvm_metrics}
    \begin{tabular}{|c|c|c|c|c|c|}
        \hline
        \textbf{Device} & \textbf{$V_\text{set}$ (V)} & \textbf{$V_\text{reset}$ (V)} & \textbf{LRS Slope} & \textbf{A} & \textbf{$A_\text{min}$} \\ \hline

        Pt/HfO$_2$ 
            & 0.778 & -0.471 & 0.0084 & $3.66\times10^{-4}$ & $6.60\times10^{-7}$ \\ \hline

        Al/Ge/TaO$_x$ 
            & 3.055 & -1.474 & 2.3032 & $1.73\times10^{-1}$ & $6.30\times10^{-4}$ \\ \hline

        Ti/SiO$_2$ 
            & 2.116 & -1.191 & 0.0049 & $7.18\times10^{-4}$ & $4.09\times10^{-7}$ \\ \hline

        Pt/HfO$_x$/TiO$_x$/TiN 
            & 1.655 & -1.404 & 0.0124 & $4.74\times10^{-2}$ & $2.97\times10^{-4}$ \\ \hline
    \end{tabular}
\end{table}

The CNN block captured the overall shape of the target I–V characteristics but showed discrepancies in key features such as $V_{\text{set}}$ and $V_{\text{reset}}$, even on the Stanford model dataset it was trained on, indicating underfitting. This arises because multiple fitting parameter sets can produce similar I–V characteristics, causing the CNN to output an averaged regression estimate. This effect is reflected in the error percentages presented in Table~\ref{tab:rram_parameters_all}, which may suggest sub-optimal performance. However, inspection of the I–V characteristics in Fig.~\ref{fig:heuristic3} demonstrates that the framework successfully reproduces the desired I-V characteristics across all devices. These observations indicate that, despite seemingly high error percentages, the framework captures the essential device physics, and that multiple parameter sets can correspond to the same I–V response, explaining the averaged output of the CNN block.
  
To improve upon the output of the CNN block, the first heuristic block improved the alignment of set and reset voltages by adjusting parameters  $\beta$ and $\gamma_{0}$ as shown in Fig.~\ref{fig:heuristic1}. Again, the second heuristic block optimized parameter $V_0$ to better match the slopes of the high-resistance state (HRS) and low-resistance state (LRS) as shown in Fig.~\ref{fig:heuristic2}. Finally, the third heuristic block tuned parameter $g_0$ to adjust the hysteresis loop area. After applying all three heuristic blocks, the I-V characteristics closely matched the target data across all key characteristics as shown in Fig.~\ref{fig:heuristic3}.

\begin{table}[!t]
    \centering
    \scriptsize
    \setlength{\tabcolsep}{3pt} 
    \caption{Comparison of reference parameters and framework-generated values for four RRAM devices.}
    \label{tab:rram_parameters_all}
    \begin{tabular}{|c|c|c|c|c|}
        \hline
        \textbf{Device} & \textbf{Parameter} & \textbf{Reference} & \textbf{Framework} & \textbf{Error (\%)} \\ \hline

        \multirow{6}{*}{Pt/HfO$_2$} 
            & $g_0$      & $2.18 \times 10^{-10}$ & $2.17 \times 10^{-10}$ & 0.41 \\ \cline{2-5}
            & $V_0$      & 0.200                   & 0.208                   & 4.06 \\ \cline{2-5}
            & $\nu$      & 10.5                    & 10.2                    & 2.86 \\ \cline{2-5}
            & $\beta$    & 2.10                    & 1.69                    & 19.7 \\ \cline{2-5}
            & $\gamma_0$ & 20.8                    & 1.41                    & 93.2 \\ \cline{2-5}
            & $I_0$      & $1.70 \times 10^{-4}$   & $2.40 \times 10^{-4}$   & 41.2 \\ \hline

        \multirow{6}{*}{Al/Ge/TaO$_x$} 
            & $g_0$      & $1.50 \times 10^{-10}$ & $1.65 \times 10^{-10}$ & 10.3 \\ \cline{2-5}
            & $V_0$      & 0.250                   & 0.261                   & 4.40 \\ \cline{2-5}
            & $\nu$      & 15                      & 9.98                    & 33.5 \\ \cline{2-5}
            & $\beta$    & 1.50                    & 1.51                    & 0.36 \\ \cline{2-5}
            & $\gamma_0$ & 12.2                    & 1.34                    & 89.0 \\ \cline{2-5}
            & $I_0$      & $1.04 \times 10^{-3}$   & $4.74 \times 10^{-4}$   & 54.4 \\ \hline

        \multirow{6}{*}{Ti/SiO$_2$} 
            & $g_0$      & $1.85 \times 10^{-10}$ & $1.94 \times 10^{-10}$ & 4.86 \\ \cline{2-5}
            & $V_0$      & 0.375                   & 0.355                   & 5.33 \\ \cline{2-5}
            & $\nu$      & $1.00 \times 10^{-9}$  & 0.0127                  & 100 \\ \cline{2-5}
            & $\beta$    & 1.80                    & 1.16                    & 35.6 \\ \cline{2-5}
            & $\gamma_0$ & 18                      & 10.8                    & 40.2 \\ \cline{2-5}
            & $I_0$      & $3.74 \times 10^{-5}$   & $2.12 \times 10^{-5}$   & 43.3 \\ \hline

        \multirow{6}{*}{Pt/HfO$_x$/TiO$_x$/TiN} 
            & $g_0$      & $2.50 \times 10^{-10}$ & $2.44 \times 10^{-10}$ & 2.36 \\ \cline{2-5}
            & $V_0$      & 0.250                   & 0.275                   & 10.0 \\ \cline{2-5}
            & $\nu$      & 10                      & 2.27                    & 77.3 \\ \cline{2-5}
            & $\beta$    & 0.800                   & 0.603                   & 24.6 \\ \cline{2-5}
            & $\gamma_0$ & 16                      & 0.551                   & 96.6 \\ \cline{2-5}
            & $I_0$      & $1.00 \times 10^{-3}$   & $1.91 \times 10^{-3}$   & 91.0 \\ \hline
    \end{tabular}
        \label{tab:rram_parameters_all}
\end{table}

\subsection{Fit with other Analytical Models:}
Next, we demonstrate the application of our fitting parameter extraction framework to two RRAM devices previously fitted using other analytical models, VTEAM model and Yakopcic model. To capture the abrupt switching behavior in such devices, a time step of $dt = 1 \times 10^{-4}$ was used in the original implementation of the Stanford RRAM model.

\subsubsection*{VTEAM Model - Pt-HfO$_2$-Ti Memristor}
\begin{figure}[!t]
    \centering
    \includegraphics[width=0.48\textwidth]{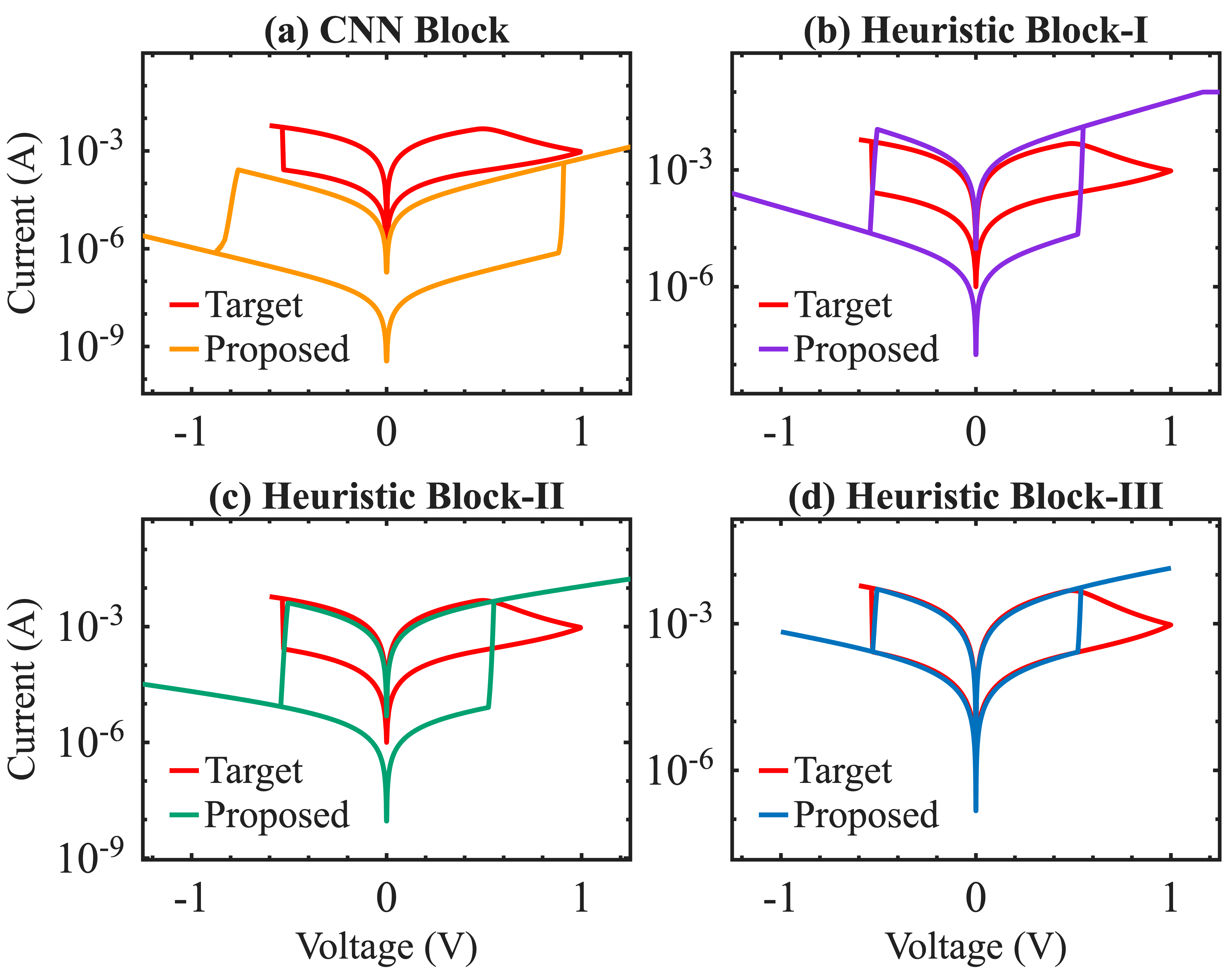} \\
    \caption{Fitting results for the VTEAM model, Pt-HfO$_2$-Ti memristor using proposed framework.(a) Initial fit obtained from the CNN block with extracted parameters: 
$g_0 = 2.39 \times 10^{-10}$, $V_0 = 0.304$, $\gamma_0 = 18.3$, 
$\beta = 0.505$, $I_0 = 1\times 10^{-4}$, and $\nu_{\text{0}} = 1.22 \times 10^{-2}$. 
Refined fits are shown for subsequent heuristic blocks: 
(b) Block~I with optimized $I_0 = 1 \times 10^{-2}$, $\gamma_0 = 0.182$, and $\beta = 0.169$; 
(c) Block~II with optimized $V_0 = 0.6$; and 
(d) Block~III with optimized $g_0 = 5 \times 10^{-10}$.}
    \label{fig:VTEAM_fitting}
\end{figure}

In this example, we model a Pt-HfO$_2$-Ti memristor using the Stanford RRAM model which was previously fitted using the VTEAM model \cite{VTeam}. Fig.~\ref{fig:VTEAM_fitting} illustrates the fitting results through the CNN and three heuristic blocks and the performance is reported in Table~\ref{tab:vteam_metrics}. The higher error percentage observed for the hysteresis area is primarily due to the gradual switching behavior exhibited by the Pt-HfO\textsubscript{2} memristor in the set region \cite{VTeam}. As VTEAM model captures the gradual set characteristics, while the Stanford RRAM model incorporates it in the reset region in its modeling. The discrepancies are thus not limitation of the proposed framework, rather intrinsic to the underlying model itself. During the fitting process, parameters $\gamma_0$, $V_0$, and $g_0$ exhibited saturation, indicating that their values had reached the imposed bounds of the parameter space. Increasing the limits to $V_{0,\max} = 0.6$, $g_{0,\max} = 8 \times 10^{-10}$, and $\gamma_{0,\max} = 30$ allowed the framework to reproduce the target I-V characteristics.

\begin{table}[!t]
    \centering
    \scriptsize
    \caption{Comparison of NVM metrics for the Pt-HfO$_2$-Ti memristor}
    \label{tab:vteam_metrics}
    \begin{tabular}{|c|c|c|c|}
        \hline
        \textbf{Metric} & \textbf{Experimental} & \textbf{Fitted} & \textbf{Error (\%)} \\ \hline
        Set Voltage ($V_\text{set}$) & 0.54 V & 0.564 V & 4.44 \\ \hline
        Reset Voltage ($V_\text{reset}$) & $-$0.55 V & $-$0.558 V & 1.45 \\ \hline
        LRS Slope & 0.01 & 0.0129 & 29.00 \\ \hline
        Hysteresis Area & 0.00384 & 0.001 & 73.95 \\ \hline
    \end{tabular}
\end{table}

\subsubsection*{Yakopcic Model - TiN/TaO\textsubscript{x}/Ta/TiN memristor}
\begin{figure}[!t]
    \centering
    \includegraphics[width=0.48\textwidth]{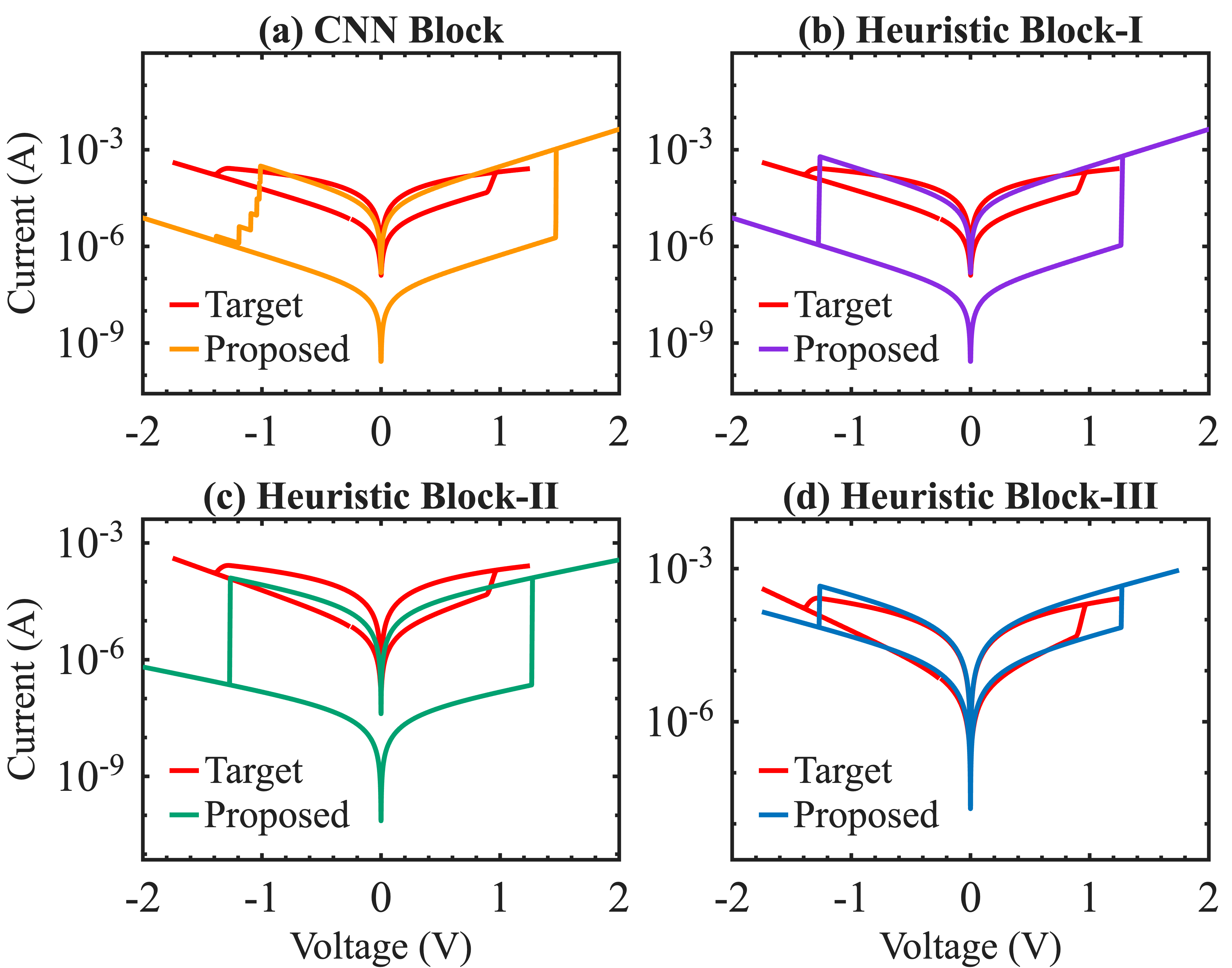}
    \caption{Fitting results for the Yakopcic model, TiN/TaO\textsubscript{x}/Ta/TiN device using proposed framework. (a) Initial fit obtained from the CNN block with extracted parameters: 
$g_0 = 2.37 \times 10^{-10}$, $V_0 = 0.377$, $\gamma_0 = 13.8$, 
$\beta = 0.867$, $I_0 = 1 \times 10^{-4}$, and $\nu_{\text{0}} = 9.39 \times 10^{-2}$. 
Refined fits are shown for subsequent heuristic blocks: 
(b) Block~I with optimized $I_0 = 2 \times 10^{-4}$, $\gamma_0 = 8.87 \times 10^{-5}$, and $\beta = 5.54 \times 10^{-5}$; 
(c) Block~II with optimized $V_0 = 0.702$; and 
(d) Block~III with optimized $g_0 = 8\times 10^{-10}$.
}
    \label{fig:yakopcic_fits}
\end{figure}

The Yakopcic model \cite{Yakopcic} simulates a TiN/TaO\textsubscript{x}/Ta/TiN memristor, with a oxide thickness ($t_\text{ox}$) set to 10 nm, consistent with the device specifications. To evaluate the ability of the Stanford RRAM model to replicate the I-V characteristics generated by the Yakopcic model, we utilized the proposed automated framework as illustrated in Fig.~\ref{fig:yakopcic_fits}.  

The generated fitting parameters yielded a sub-optimal fit for the device characteristics modeled in Stanford RRAM model as presented in Table~\ref{tab:yakopcic_metrics}. Upon inspection, the I-V characteristics derived from the Yakopcic model exhibits significant differences from the ones used to train the CNN block. As a result, the initial fit obtained from the CNN block was poor, as anticipated. Furthermore, the first heuristic block also failed to satisfy the set voltage criterion. Upon further analysis, the $\beta$ parameter generated by the framework was $8 \times 10^{-5}$, suggesting that the heuristic attempted to optimize the parameters but was limited by the initial configuration of the CNN block and constraints imposed on the heuristic block to ensure numerical stability. Fitting parameters $V_0$ and $g_0$ also exhibited saturation, and their upper bounds were increased to $V_{0,\max} = 0.8$ and $g_{0,\max} = 8 \times 10^{-10}$, similar to the adjustment made earlier for the VTEAM model.

After one complete loop through the designed framework, it was observed that the LRS slope could be further adjusted to achieve a better fit. The second heuristic block was invoked again with the fitting parameters generated by the framework upon one complete pass, resulting in an improved fit compared to the initial result.

\begin{figure}[!t]
    \centering
    \includegraphics[width=0.5\textwidth]{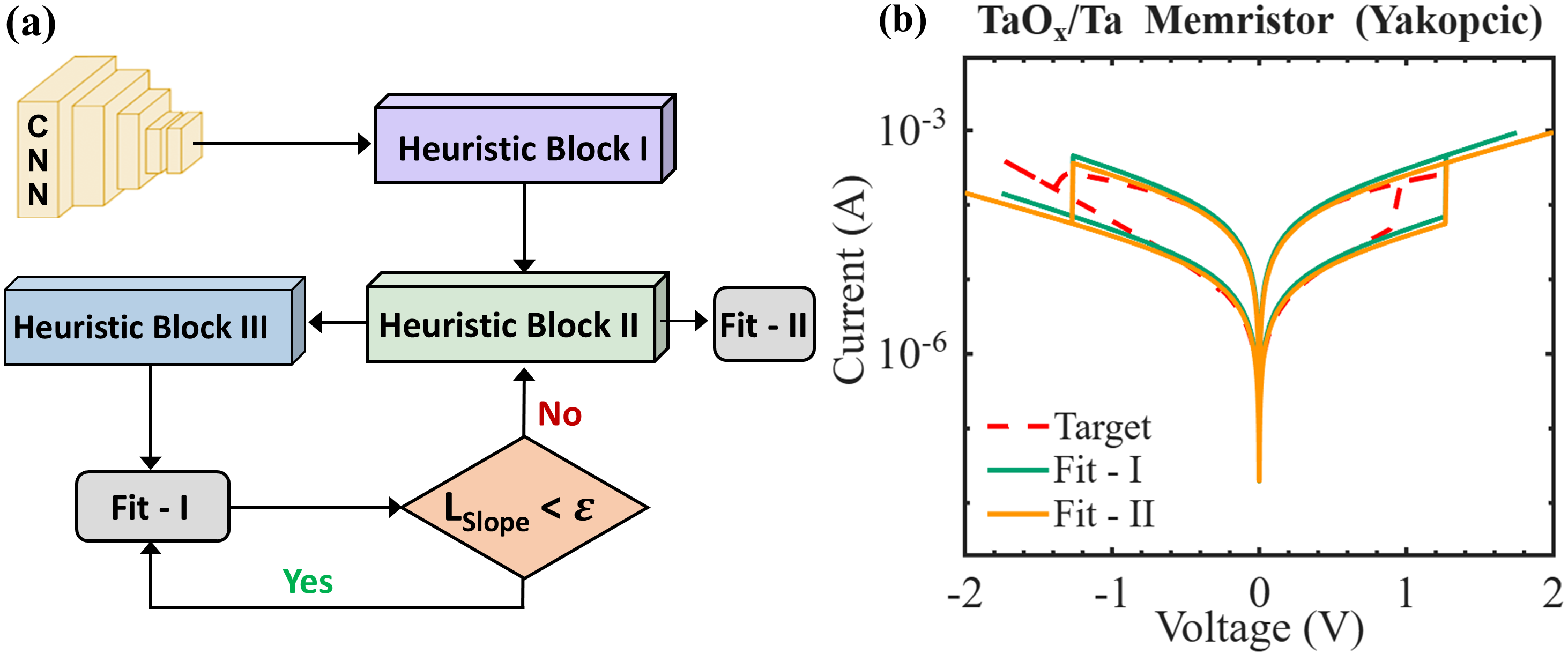}
    \caption{Refined fit obtained after reinvoking Heuristic Block-II for further slope adjustment. (a) The schematic illustrates the proposed framework's iterative process, where Heuristic Block-II is re-invoked after one complete sequence. (b) The resulting refined LRS slope after re-invoking Heuristic Block-II with optimized $V_o$ = 0.8}
    \label{fig:heuristic_block_2_refined}
\end{figure}

\begin{table}[!t]
    \centering
    \scriptsize
    \caption{Comparison of NVM metrics for the TiN/TaO\textsubscript{x}/Ta/TiN device fitting}
    \label{tab:yakopcic_metrics}
    \begin{tabular}{|c|c|c|c|}
        \hline
        \textbf{Metric} & \textbf{Experimental} & \textbf{Fitted} & \textbf{Error (\%)} \\
        \hline
        Set Voltage ($V_\text{set}$) & 0.88 V & 1.276 V & 45.00 \\
        \hline
        Reset Voltage ($V_\text{reset}$) & $-$1.28 V & $-$1.276 V & 0.31 \\
        \hline
        LRS Slope & $1.9 \times 10^{-4}$ & $2.1 \times 10^{-4}$ & 10.50 \\
        \hline
        Hysteresis Area & $2 \times 10^{-4}$ & $3 \times 10^{-4}$ & 50.00 \\
        \hline
    \end{tabular}
\end{table}

\subsection{Fit with Experimental Data}

Lastly, we showcase two examples of fitting parameter extraction performed on I-V characteristics extracted from the literature. To prepare the data collected from literature for our framework, the I-V characteristics were extracted using WebPlotDigitizer, followed by pre-processing using a rolling window averaging technique to reduce noise. Additionally, the compliance current in the framework had to be set accordingly to accurately fit these I-V characteristics, as it significantly influenced the current-voltage relationship observed in the experimental data.

\subsubsection*{Fitting TiN/HfO$_\mathbf{x}$/AlO$_\mathbf{x}$/Pt Memristor}

\begin{figure}[!t]
    \centering
   \includegraphics[width=0.48\textwidth]{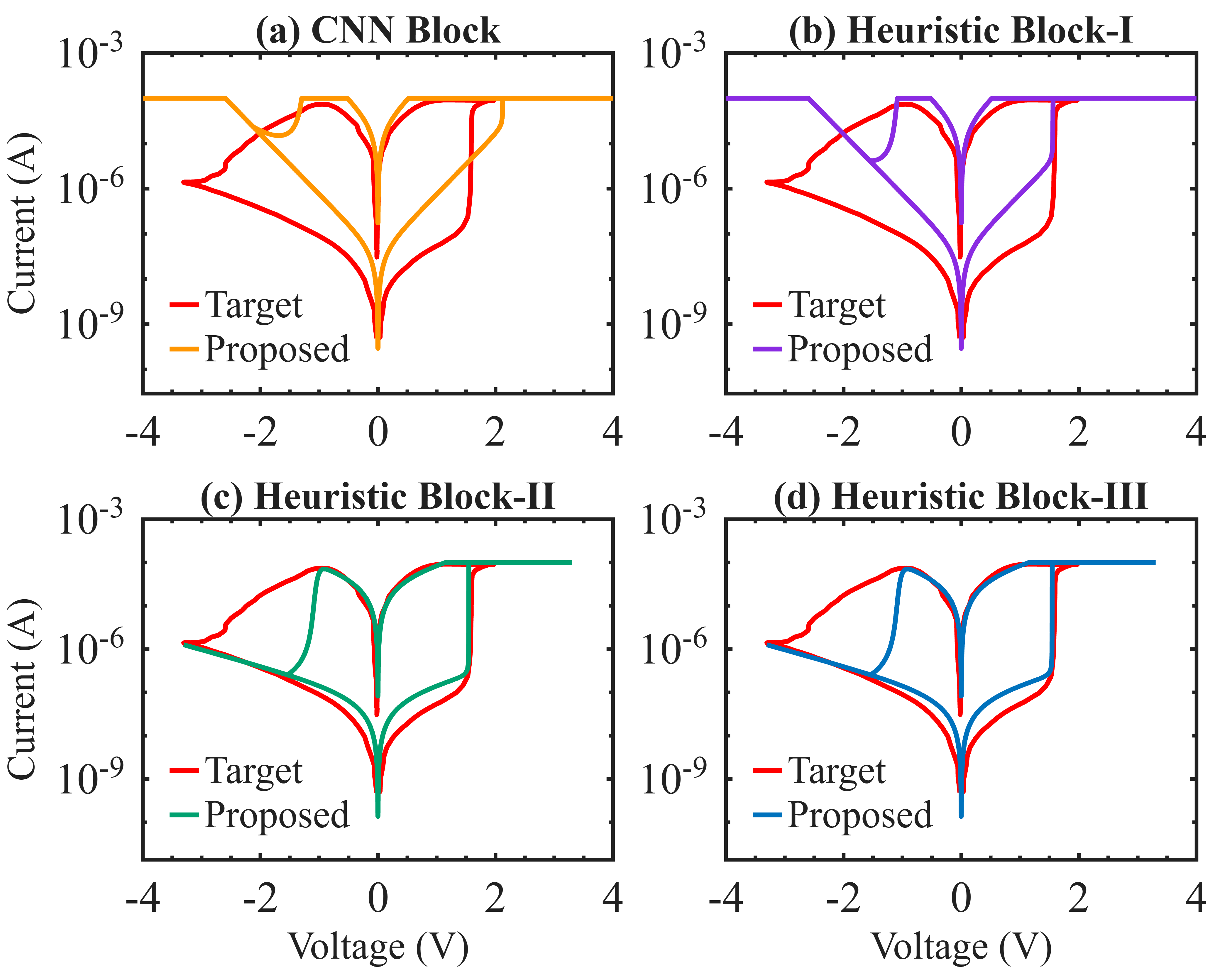}
    \caption{Fitting results for the TiN/HfO$_x$/AlO$_x$/Pt memristor using proposed framework. 
(a) Initial fit obtained from the CNN block with extracted parameters: 
$g_0 = 2.35 \times 10^{-10}$, $V_0 = 0.327$, $\gamma_0 = 18.6$, 
$\beta = 1.85$, $I_0 = 1 \times 10^{-4}$, and $\nu_{\text{0}} = 6.75 \times 10^{-2}$. 
Refined fits are shown for subsequent heuristic blocks: 
(b) Block~I with optimized $I_0 = 2 \times 10^{-4}$, $\gamma_0 = 21.9$, and $\beta = 1.85$; 
(c) Block~II with optimized $V_0 = 0.8$; and 
(d) Block~III with optimized $g_0 = 2 \times 10^{-10}$. 
    \label{fig:TiN_fitting} }
\end{figure}

We demonstrate the application of the automated framework in fitting the I-V characteristics of a TiN/HfO$_x$/AlO$_x$/Pt memristor extracted from \cite{Ti}, as illustrated in figure~\ref{fig:TiN_fitting}. The compliance current was set to $1\times10^{-4}$~A as reported in the paper. 

 In this case, parameters generated by the CNN block provided a strong initial estimate, significantly reducing the need for further corrections. Upper bound for $V_o$ was updated to $0.8$ for optimal LRS slope matching. The quality of proposed fit is summarized in Table~\ref{tab:TiN_metrics}. The significant error in the hysteresis area is attributed to the framework’s inability to model the gradual reset process, a characteristic of the TiN/HfO$_x$/AlO$_x$/Pt memristor observed in the experimental data. Such artifact can be introduced in the Stanford model as well with inclusion of an extra parameter \cite{Stanford}.

\begin{table}[!t]
    \centering
    \scriptsize
    \caption{Comparison of NVM metrics for the TiN/HfO$_x$/AlO$_x$/Pt memristor.}
    \label{tab:TiN_metrics}
    \begin{tabular}{|c|c|c|c|}
        \hline
        \textbf{Metric} & \textbf{Experimental} & \textbf{Fitted} & \textbf{Error (\%)} \\ \hline
        Set Voltage ($V_\text{set}$) & $1.55$ V& $1.551$ V& $0.06$ \\ \hline
        Reset Voltage ($V_\text{reset}$) & $-1.1$ V& $-1.097$ V& $0.27$ \\ \hline
        LRS Slope & $1\times10^{-4}$ & $1.1\times10^{-4}$ & $10.00$ \\ \hline
        Hysteresis Area & $2.22\times10^{-4}$ & $1.11\times10^{-4}$ & $50.00$ \\ \hline
    \end{tabular}
\end{table}

\subsubsection*{Fitting Pt/TiO$_\mathbf{2}$ NR Array/FTO Device}
\begin{figure}[!t]
    \centering
    
       \includegraphics[width=0.48\textwidth]{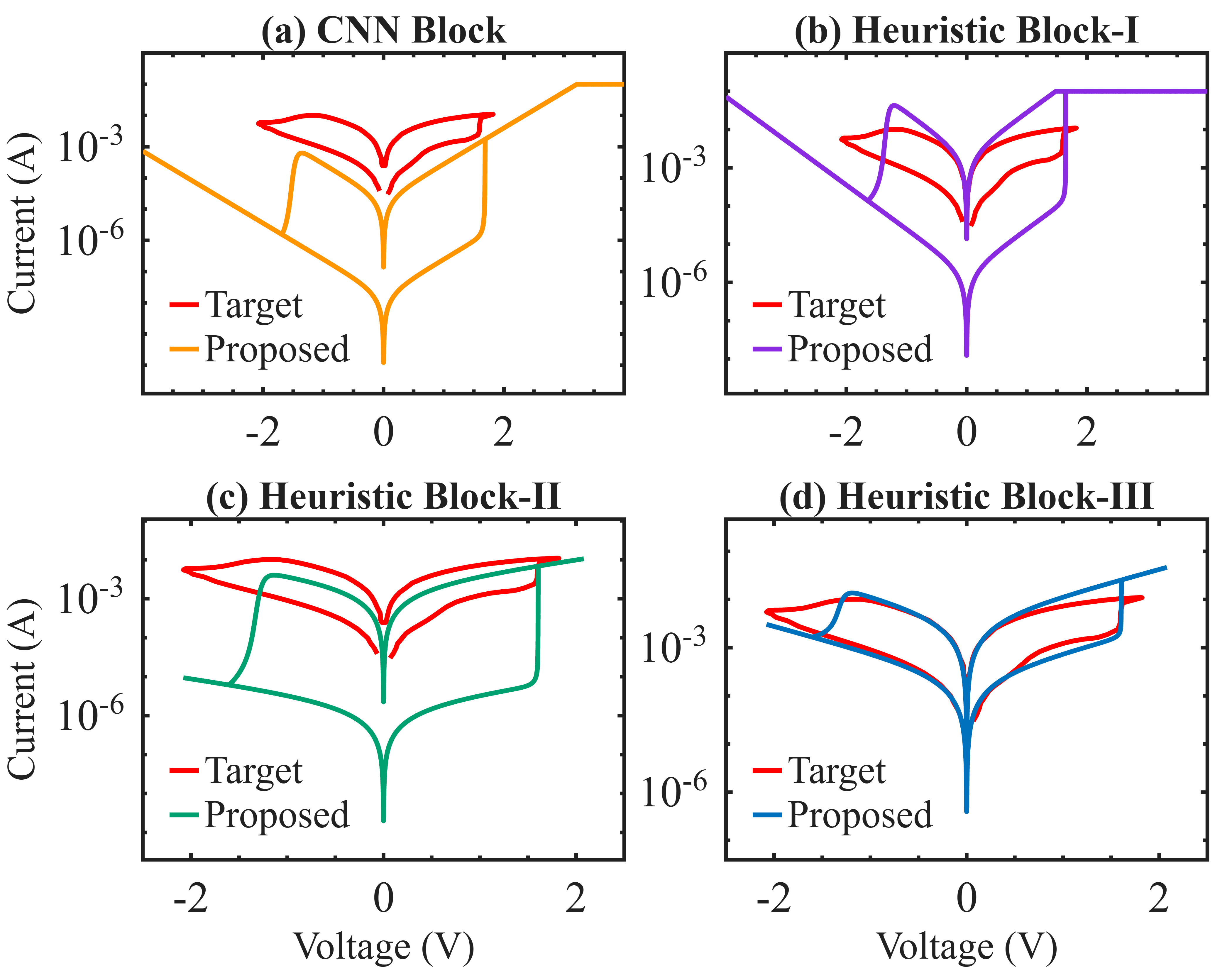} 
    \caption{Fitting results for the Pt/TiO$_2$ NR array/FTO device using proposed framework. 
(a) Initial fit obtained from the CNN block with extracted parameters: 
$g_0 = 2.14 \times 10^{-10}$, $V_0 = 0.376$, $\gamma_0 = 19.5$, 
$\beta = 0.59$, $I_0 = 1\times 10^{-4}$, and $\nu_{\text{0}} = 2.76 \times 10^{-4}$. 
Refined fits are shown for subsequent heuristic blocks: 
(b) Block~I with optimized $I_0 = 1.5 \times 10^{-2}$, $\gamma_0 = 1.01$, and $\beta = 1.09$; 
(c) Block~II with optimized $V_0 = 1.21$; and 
(d) Block~III with optimized $g_0 = 6.18 \times 10^{-10}$.}

    \label{fig:Pt_TiO2_Stages}
\end{figure}

The application of the automated framework for fitting the I-V characteristics of Pt/$TiO_2$ NR array/FTO devices \cite{NR} is illustrated in Fig. ~\ref{fig:Pt_TiO2_Stages} and the quality of the fit is summarized in Table~\ref{tab:Pt_TiO2_metrics}. In this case, the CNN block generated an initial fit that was significantly off from the target I-V characteristics similar to Fig. ~\ref{fig:yakopcic_fits}. Upper bounds for $V_0$ and $g_0$ were set to $1.5$ and $6.18 \times 10^{-10}$ respectively for optimal fitting. Also, due to the asymmetric slopes in the set and reset regions, the reset region fit was prioritized, leading to a slight mismatch in the set region. 

\begin{table}[!t]
    \centering
    \scriptsize
    \caption{Comparison of NVM metrics for the Pt/TiO\textsubscript{2} NR Array/FTO Device}
    \label{tab:Pt_TiO2_metrics}
    \begin{tabular}{|c|c|c|c|}
        \hline
        \textbf{Metric} & \textbf{Experimental} & \textbf{Fitted} & \textbf{Error (\%)} \\
        \hline
        Set Voltage ($V_\text{set}$) & 1.65 V & 1.647 V & 0.18 \\
        \hline
        Reset Voltage ($V_\text{reset}$) & $-$1.35 V & $-$1.337 V & 0.96 \\
        \hline
        LRS Slope & $6.7 \times 10^{-3}$ & $6.73 \times 10^{-3}$ & 0.45 \\
        \hline
        Hysteresis Area & $2.4 \times 10^{-2}$ & $7 \times 10^{-3}$ & 70.83 \\
        \hline
    \end{tabular}
\end{table}

The proposed automated fitting parameter extraction framework was benchmarked across various RRAM device I–V characteristics, spanning both previously fitted ones with Stanford RRAM model used during training and entirely unseen experimental device characteristics reported in the literature. In all cases, the framework successfully generated fitting parameters that enabled reasonable reproduction of experimental I–V behavior using the Stanford RRAM model. Quantitative evaluation of the error percentages in the proposed reference NVM metrics further confirmed the effectiveness of the framework, indicating its capability to achieve optimal performance. However, per metric analysis in some cases indicated otherwise due to a significant error percentage, these deviations were primarily attributed to inherent limitations of the Stanford RRAM model in capturing device specific artifacts commonly observed in RRAM technologies. Such unique artifacts such as synaptic behaviour or asymmetric set and reset slope etc. are not the primary concern of circuit designers focusing on NVM applications. Consequently, the proposed framework offers a significant advantage over manual parameter tuning by enabling faster extraction of fitting parameters while maintaining comparable quality in modeling RRAM I–V characteristics.

\section{Conclusion}
In this work, we introduced an automated fitting parameter extraction framework for the Stanford RRAM model, combining convolutional neural network with three heuristic blocks each of which adaptively optimize for a specific NVM reference metric. The framework is designed to be user-friendly, requiring only the device I–V characteristics and a few directly extractable characteristic quantities for producing accurate fitting results. It has demonstrated good performance in fitting a wide range of RRAM I–V characteristics, including devices not previously modeled with the Stanford model, such as those described by the VTEAM and Yakopcic models, as well as other experimentally reported devices. Beyond the Stanford model, the framework can be adapted to automate parameter extraction for other compact models as well. By streamlining the device modeling process, the proposed framework bridges the gap between device-level physics and circuit design practice, and has the potential to significantly accelerate the development and application of RRAM technologies.

\singlespacing
\bibliographystyle{IEEEtran}
\bibliography{buetug}

\end{document}